\documentclass[10pt,conference]{IEEEtran}
\usepackage{amsfonts}
\usepackage{dsfont}
\usepackage{setspace}

\usepackage{setspace}
\usepackage{amsmath}
\usepackage{amssymb}
\usepackage[dvips]{graphicx}
\usepackage{epsfig}
\usepackage{dsfont}

\usepackage[ps2pdf,
bookmarks=false,
bookmarksnumbered=false, 
bookmarksopen=false, 
colorlinks=false]{}

\newcommand{\E}{\mathbb{E}}

\newcommand{\BS}{{\mathbf{BS}}}

\begin{document}

\title{A Delay-Aware Caching Algorithm for Wireless D2D Caching Networks}

\author{Yi Li, M. Cenk Gursoy and Senem Velipasalar
\\Department of Electrical Engineering and Computer Science,
Syracuse University, Syracuse, NY 13244
\\Email: yli33@syr.edu, mcgursoy@syr.edu, svelipas@syr.edu}

\maketitle

\begin{abstract} \let\thefootnote\relax\footnote{This work was supported in part by National Science Foundation grants CCF-1618615, ECCS-1443994, CNS-1302559, and CNS-1206291.}
Recently, wireless caching techniques have been studied to satisfy lower delay requirements and offload traffic from peak periods. By storing parts of the popular files at the mobile users, users can locate some of their requested files in their own caches or the caches at their neighbors. In the latter case, when a user receives files from its neighbors, device-to-device (D2D) communication is enabled. D2D communication underlaid with cellular networks is also a new paradigm for the upcoming 5G wireless systems. By allowing a pair of adjacent D2D users to communicate directly, D2D communication can achieve higher throughput, better energy efficiency and lower traffic delay. In this work, we propose a very efficient caching algorithm for D2D-enabled cellular networks to minimize the average transmission delay. Instead of searching over all possible solutions, our algorithm finds out the best $<$file,user$>$ pairs, which provide the best delay improvement in each loop to form a caching policy with very low transmission delay and high throughput. This algorithm is also extended to address a more general scenario, in which the distributions of fading coefficients and values of system parameters potentially change over time. Via numerical results, the superiority of the proposed algorithm is verified by comparing it with a naive algorithm, in which all users simply cache their favorite files.
\end{abstract}

\section{Introduction}
Recently, many studies have been conducted to analyze caching strategies in wireless networks in order to satisfy the throughput, energy efficiency and latency requirements in next-generation 5G wireless systems. By storing parts of the popular files at the base station and users' devices, network traffic load can be managed/balanced effectively, and traffic delay can be greatly reduced. It has been pointed out that $60\%$ of the content is cacheable in the network traffic \cite{Cisco}, which can be transmitted and stored close to the users before receiving the requests. A brief overview of wireless caching was provided in \cite{WL_caching_overview}, which introduced the key notions, challenges, and research topics in this area. In order to improve the performance effectively, the system needs to estimate and track the popularity of those cacheable contents, and predict the popularity variations, helping to guarantee that the most popular contents are cached and the outdated contents are removed. In \cite{LivingOnTheEdge}, popularity matrix estimation algorithms were studied for wireless networks with proactive caching.

Multiple caching strategies have been investigated in the literature, which improve the performance in different ways. When contents are cached at the base stations, the energy consumption, traffic load and delay of the backhaul can be reduced \cite{ClusterContentCaching}, and the base stations in different cells can cooperate to improve the spectral efficiency gain \cite{PhyCaching5G}. When contents are cached at the users' devices, the base station can combine different files together and multicast to multiple users, and the users can decode their desired files using their cached files. A content distribution algorithm for this approach was given in \cite{MulticastingCodingRandomDemands}, and the analysis of the coded multicasting gain was provided in \cite{CodingforCaching}.

D2D communication underlaid with cellular networks is another technology that has attracted much interest recently. In D2D communication, users can communicate directly without going through the base station. The advantages of D2D communications were studied in \cite{D2D_KB}, and it was shown that D2D communication could greatly enhance the spectral efficiency and lower the latency. A comprehensive overview was provided in \cite{D2D_survey}, where different modeling assumptions and key considerations in D2D communications were detailed. In a D2D cellular network, users can choose to work in different modes. In cellular mode, users communicate through the base station just as cellular users; while in D2D mode, users communicate directly. Mode selection is a critical consideration in D2D communications, and many studies have been conducted in this area. For example, in \cite{mode_selection_DK}, mode selection problem was studied for a system with one D2D pair and one cellular user, and in \cite{Joint_selection}, a joint mode selection and resource allocation algorithm was proposed. Recently in \cite{li2016device}, mode selection and optimal resource allocation in D2D networks were studied under statistical queueing constraints.

In the literature, several studies have been performed to combine content caching with device-to-device (D2D) wireless networks. In such cases a user can receive from its neighbors if these have cached the requested content. An overview on wireless D2D caching networks was provided in \cite{WirelessD2DCachingNetworks}, in which the key results for different D2D caching strategies were presented. To design caching policies for the wireless D2D network, the authors of \cite{MobileCachingD2DContentDelivery} proposed a caching policy that maximizes the probability that requests can be served via D2D communications. For a similar system setting, a caching policy that maximizes the average number of active D2D links was obtained in \cite{golrezaei2012wireless}. Most of these works were based on stochastic geometry models, in which nodes/users were distributed randomly. However, these types of models mainly focus on the path loss, and do not fully address the effects of channel fading. Without the characterization of the channel fading, an accurate analysis on the throughput and delay is not viable. Moreover, many works only tackle a simple case in which users have identical popularity vectors. In this paper, we design a caching algorithm that minimizes the average delay of the network, and our main contributions are listed as follows:
\begin{enumerate}
\item We provide a characterization of the average delay in both cellular and D2D modes.
\item Our algorithm minimizes the average delay of the system, which is a significant objective in real-time applications.
\item We propose a very efficient and robust algorithm to solve the delay minimization problem.
\item Our algorithm is applicable in settings with very general popularity models, with no assumptions on how file popularity varies among different users.
\item We further extend our algorithm to a more general setting, in which the system parameters and the distributions of channel fading change over time.
\end{enumerate}

\section{System Model and Problem Formulation}\label{Sec:formulation}

\subsection{System Model and Channel Allocation} \label{sec:model}
\begin{figure}
\begin{center}
\includegraphics[width=0.4\textwidth]{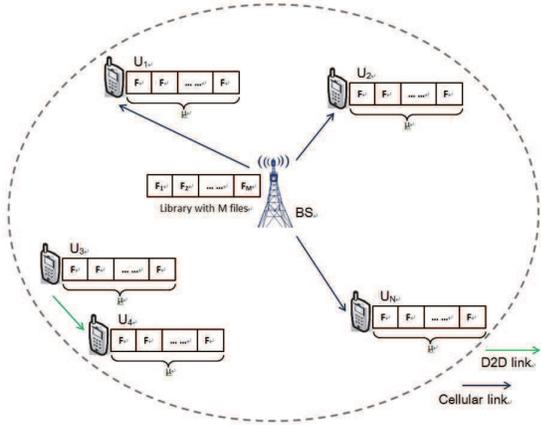}
\caption{System model of a D2D cellular network with caches}\label{fig:system}
\end{center}
\end{figure}

As shown in Figure \ref{fig:system}, we consider a cellular network with one base station ($\BS$), in which a library with $M$ files ($\mathcal{F}_1$, $\mathcal{F}_2$, $\cdots$, $\mathcal{F}_M$) is stored, and we assume that the size of each file is fixed to $F$ bits \footnote{In the literature, it is noted that the base station may only store a portion of the library contents, and needs to acquire the remaining files from the content server \cite{WL_caching_overview}. Since we focus on the wireless transmission delay, we do not explicitly address the link between the base station and content server. Also, the content files may not have the same size in practice, but we can further divide them into sub-files with equal size.}. There are $N$ users ($U_1$, $U_2$, $\cdots$, $U_N$) in the network who seek to get the content files from the library. Each user is equipped with a cache of size $\mu F$ bits, and therefore can store $\mu$ content files. The caching state is described by an $N\times M$ matrix $\mathbf{\Phi}$, whose $(i,j)$-th component has a value of $\phi_{i,j}=1$ if file $\mathcal{F}_j$ is cached at user $U_i$, and $\phi_{i,j}=0$ when the user $U_i$ does not have file $\mathcal{F}_j$ in its cache.

In general, users request files with different probabilities, which are characterized by an $N\times M$ popularity matrix $\bold{P}$, in which the entry on the $i^{\text{th}}$ row and $j^{\text{th}}$ column, $P_{i,j}$, represents the probability of user $U_i$ requesting file $\mathcal{F}_j$. Each row of the popularity matrix corresponds to a popularity vector of a user. Although the popularity matrix may change over time in practice, we can assume that the popularity stays constant within a certain period, and our caching algorithm needs to be repeated when the popularity matrix is updated. In the literature, Zipf distribution is generally considered as a good statistical model for the popularity. The probability mass function (pmf) of this distribution is given by
\begin{align}\label{eq:Zipf_pmf}
P_{i,j}=\frac{f_{i,j}^{-\beta}}{\sum_{k=1}^{M} k^{-\beta}},
\end{align}
where $f_{i,j}$ is the popularity index that user $U_i$ gives to file $\mathcal{F}_j$, and $\beta \geq 0$ is the Zipf exponent. Each user enumerates the files with popularity index from $1$ to $M$, where the most popular file gets index $1$, and the least popular file gets index $M$. As the Zipf exponent $\beta$ increases, the difference in the popularity of different files increases, while all files have the same popularity when $\beta\rightarrow 0$. Although we use Zipf distribution for our numerical results in Section \ref{Sec:numerical}, our proposed algorithm works for any type of popularity model. At each user, the generated requests are buffered in a queue before getting served, and it is assumed that these request queues are not empty at any time.

In a D2D-enabled wireless network, users can choose to transmit in cellular mode or D2D mode. In the cellular mode, users request and receive information from the base station, while in the D2D mode, a user requests and receives information from another user through a D2D direct link. In our model, the users first check their local cache when a file is requested. If the user does not have the corresponding file in its own cache, it sends a request to the base station. We assume that the base station has knowledge of all fading \emph{distributions} (i.e., only has statistical information regarding the channels) and the cached files at each user. After receiving the request, the base station identifies the source node from which the file request can be served and allocates channel resources to the corresponding user. Therefore, the result of mode selection is determined by the result of source selection. If the source node is another user, then the requested file is sent over the direct D2D link and hence the communication is in D2D mode, otherwise the receiving user works in cellular mode and receives files from the base station. In source selection, among all the nodes (including the base station) who have the requested file, the node with the lowest average transmission delay to the receiver is selected as the transmitter.

In this work, we consider an OFDMA system with $N_c$ orthogonal channels, and the bandwidth of each channel is $B$. We assume that the background noise samples follow independent and identically distributed (i.i.d.) circularly-symmetric complex Gaussian distribution with zero mean and variance $\sigma^2$ at all receivers in all frequency bands, and the fading coefficients of the same transmission link are i.i.d. in different frequency bands. The fading coefficients are assumed to stay constant within one time block of duration $T_0$, and change across different time blocks. We summarize the resource allocation assumptions for the discussions in Sections \ref{Sec:formulation} and \ref{Sec:algorithm} as follows:
\begin{enumerate}
  \item Each channel can be used for the transmission of one requested file at most, and the transmission of a file cannot occupy multiple channels.
  \item D2D transmitters are not allowed to transmit to multiple receivers simultaneously. In other words, the file requests whose best source node is a user who is already transmitting cannot be assigned a channel resource by the base station.
  \item The probability of a channel being allocated to a request generated by user $i$ is $\hat{p}_{i}$.
  \item After a request is served, the corresponding transmitter keeps silent in the remaining time block, and the base station allocates the channel resource to other requests at the beginning of the following time block.
  \item If the $i^{\text{th}}$ user is selected as a D2D transmitter, its maximum transmission power is $P_i$.
  \item The base station can serve multiple requests simultaneously using different channels, and its maximum transmission power is $P_b$ for each request.
\end{enumerate}

The first four assumptions describe a class of simple scheduling algorithms, in which only point to point transmission without spectrum reusing is considered. At the beginning of each time block, base station assigns available channels to the requests, and each transmission link gets one channel at most, and uses the assigned channel exclusively. The transmitter transmits until the request is served and then releases the channel resource. The behavior of the scheduling algorithm is described by a set of probabilities $\hat{p}_{i}$ defined in the third assumption. Although our delay characterizations in Sections \ref{Sec:delay} and \ref{Sec:problem} are only valid for this type of scheduling algorithms, we further extend our results for more complicated scheduling algorithms in Section \ref{sec:extension}. In that case, only the last two assumptions are required, which describe the maximum power constraints. With a more complicated scheduling algorithm, we can only estimate the average delay of each request at each user through simulation or learning methods. A detailed discussion is provided in Section \ref{sec:extension}.

In this work, only the distributions of the fading coefficients are required at the base station, which mainly depend on the environment and the location of each user. A centralized computation scheme is used, and the base station sends the results of caching and scheduling algorithms to the users through additional control channels. Since the base station knows the distributions of all fading coefficients and the cached files at each user, the average delay between each user pair and the best source node for each request can be obtained at the beginning and stored in tables at the base station. A detailed discussion on delay calculation and the determination of the best source node is provided in the next subsection.

\subsection{Transmission Delay}\label{Sec:delay}
In this work, we use the transmission delay, which is defined as the number of time blocks used to transmit a content file, as the performance metric. From the above discussion, the instantaneous channel capacity a transmission link in the $k^{\text{th}}$ time block is
\begin{small}
\begin{align}
C[k]=B\log_2\left(1+\frac{P_t}{B\sigma^2}z_k\right)\hspace{0.3cm} \text{bits}/\text{s}
\end{align}
\end{small}
where $P_t$ is the transmission power, and $z_i$ is the magnitude square of the corresponding fading coefficient in the $k^{\text{th}}$ time block. In order to maximize the transmission rate, all transmitters transmit at the maximum power level. Therefore,
\begin{small}
\begin{align}
P_t=
\begin{cases}
P_b\hspace{1cm} \text{if the transmitter is the base station}\\
P_i\hspace{1cm} \text{if the transmitter is the $i^{\text{th}}$ user}
\end{cases},
\end{align}
\end{small}
and the duration to send a file is
\begin{small}
\begin{align}
T=\min\left\{t:F\leq\sum_{k=1}^{t} T_0 C[k]\right\}
\end{align}
\end{small}
where $F$ is the size of each file, $T_0$ is the duration of each block, and $C[k]$ is the instantaneous channel capacity in the $k^{\text{th}}$ time block. When the fading distribution is available, the average transmission delay of the link $U_i-U_j$, which is denoted by $\E\{T_{i,j}\}$, can be obtained through numerical methods or Monte-Carlo simulations. These average delay values can be stored in an $N\times N$ symmetric matrix $\bold{T_{\text{avg}}}$, whose component on the $i^{\text{th}}$ row and $j^{\text{th}}$ column is given by $\E\{T_{i,j}\}$ when $i\neq j$, and the diagonal element $T_{i,i}$ is the average delay between $U_i$ and the base station. According to our channel assumptions, the average delays of a transmission link are the same in every channel. Therefore, we only need to analyze the performance in a single channel.

The best source node of the request, which is generated by user $U_i$ requesting file $\mathcal{F}_j$, is the node which has file $\mathcal{F}_j$ and the smallest average transmission delay to $U_i$, and this minimum average delay is denoted by $D_{i,j}$ \footnote{If $U_i$ has cached $\mathcal{F}_j$, then the best source node is $U_i$ itself, and $D_{i,j}=0$.}. The best source of each possible request can be stored in an $N\times M$ table $\bold{S}$, in which each row corresponds to a user who generates the request, and each column corresponds to a file being requested. Also, these $D_{i,j}$ values can be collected in an $N\times M$ matrix $\bold{D}$.

Using the above results, the average transmission delay of the requests generated by user $U_i$ can be obtained as
\begin{small}
\begin{align}
D_i=\sum_{j=1}^{M} P_{i,j}D_{i,j},
\end{align}
\end{small}
where $P_{i,j}$ is the $(i,j)$-th component of the popularity matrix $\bold{P}$.

\subsection{Problem Formulation}\label{Sec:problem}
In the previous subsection, we have determined and expressed the average delay. In this subsection, we formulate and discuss our caching problem. In this work, our goal is to minimize the weighted sum of the average delays of the users, which is expressed as
\begin{small}
\begin{align}
\eta &=\sum_{i=1}^{N} \omega_i D_i =\sum_{i=1}^{N} \omega_i \sum_{j=1}^{M} P_{i,j}D_{i,j} \label{eq:eta}
\end{align}
\end{small}
where $\omega_i \in [0,1]$ is the weight for user $U_i$. We assume that the values of the weights are predetermined. In practice, $\omega$ values can be determined according to the priorities of users, so that users with higher priority have higher weights.

Our caching problem is formulated as
\begin{small}
\begin{align}
\textbf{P1:}\hspace{1cm}&\text{Minimize}_{\;\mathbf{\Phi}}\hspace{1.2cm} \eta\\
&\text{Subject to} \hspace{0.7cm} \sum_{j=1}^{M} \phi_{i,j}=\mu \label{constraint_1}\\
&\hspace{2.2cm} \phi_{i,j}\in\{0,1\}
\end{align}
\end{small}
where $\mathbf{\Phi}$ is the caching result indicator matrix. The constraint in (\ref{constraint_1}) arises due to the maximum cache size. It is obvious that the optimal caching policy must use all caching space.

In a special case, if we choose $\omega_i=\hat{p}_{i}$, where $\hat{p}_{i}$ is the probability that a channel is allocated to user $U_i$, then $\eta$ expresses the average delay of the system. In this situation, the throughput of the system can be expressed as
\begin{small}
\begin{equation}\label{eq:throughput}
R=N_c\frac{F}{\eta}.
\end{equation}
\end{small}
Therefore, in this special case, minimizing $\eta$ is equivalent to maximizing the throughput of the system.

\section{Caching Algorithm}\label{Sec:algorithm}
In this section, we propose our caching algorithm that solves problem $\textbf{P1}$. Note that the objective in problem $\textbf{P1}$ is not convex, and the solution space is a discrete set with size $(\frac{M!}{(M-\mu)!\,\mu!})^N$. Therefore, the globally optimal solution can only be obtained via exhaustive search. In this work, we propose an efficient algorithm to determine a caching policy with delay performance close to the optimal solution. At the end of this section, we show that our algorithm has the potential to be extended to more complicated scenarios.

\subsection{Caching Algorithm}
Our algorithm is a greedy algorithm, which searches over a subset of the solution space with smaller size. At the beginning, we assume that all caches are empty, and every user has to operate in cellular mode, in which they only receive files from the base station. Then, in each step, we find the best $<$file,user$>$ pair, which provides the maximum delay improvement (or equivalently reduction in delay) if the selected file is stored in the cache of the corresponding user. This process needs to be repeated $N\mu$ times, in order to fill all cache space, and the final caching policy is obtained.

\begin{table}
\caption{\label{Algorithm1}Algorithm 1}
\begin{footnotesize}
\begin{tabular}{p{8cm}}
\hline
\hline
Find the delay improvement for a $<$file,user$>$ pair\\
\hline
\hline
\textbf{Input :} user index $i$, file index $j$, caching indicator $\phi_{i,j}$, weight vector $\boldsymbol{\omega}=(\omega_1,\cdots,\omega_N)$, popularity matrix $\bold{P}$, source table $\bold{S}$, delay matrices $\bold{T_{\text{avg}}}$ and $\bold{D}$.\\
\textbf{Output :} delay improvement $g_{i,j}$, updated source table $\hat{\bold{S}}$, updated optimal delay matrix $\hat{\bold{D}}$.\\
\hline
\textbf{Initialization :} $\hat{\bold{S}}=\bold{S}$ and $\hat{\bold{D}}=\bold{D}$\\

\textbf{If} $\phi_{i,j}=1$\\
\hspace{0.5cm} $g_{i,j}=0$, end process.\\
\textbf{Else}\\
\hspace{0.5cm} $g_{i,j}=\omega_i P_{i,j} D_{i,j}$ and update $\hat{S}_{i,j}\leftarrow U_i$, $\hat{D}_{i,j}=0$.\\
\textbf{End}\\
\\
\textbf{For} $k=1:N$\\
\hspace{0.5cm} \textbf{If} $D_{k,j}>T_{i,k}$ and $i\neq k$\\
\hspace{1cm} $g_{i,j}=g_{i,j}+\omega_{k} P_{k,j} (D_{k,j}-T_{i,k})$ \\
\hspace{1cm} update $\hat{D}_{k,j}=T_{i,k}$ and $\hat{S}_{k,j}\leftarrow U_i$\\
\hspace{0.5cm} \textbf{End}\\
\textbf{End}
\\
\hline\hline
\end{tabular}
\end{footnotesize}
\end{table}

In Table \ref{Algorithm1}, we describe Algorithm 1 in detail, which calculates the delay improvement and determines the updated $\bold{S}$ and $\bold{D}$ matrices accordingly when we cache file $\mathcal{F}_j$ at user $U_i$. First, we check if $\mathcal{F}_j$ has already been cached at $U_i$. If so, we end the process, and return the delay improvement $g_{i,j}=0$; if not, we set $g_{i,j}=\omega_i P_{i,j} D_{i,j}$ because that is the reduction in $\eta$ at user $U_i$ if it adds $\mathcal{F}_j$ to its cache. Then, we need to sum up all reductions at each user. At user $U_k$, if $D_{k,j}>T_{i,k}$, then D2D link $U_i-U_k$ has the lowest average delay for $U_k$ to receive $\mathcal{F}_j$ and the reduction at $U_k$ is $\omega_{k} P_{k,j} (D_{k,j}-T_{i,k})$; if not, then caching $\mathcal{F}_j$ at $U_i$ does not help to improve the delay performance at $U_k$.

\begin{table}
\caption{\label{Algorithm2}Algorithm 2}
\begin{footnotesize}
\begin{tabular}{p{8cm}}
\hline
\hline
Find the optimal $<$file,user$>$ pair to be added in the updated caching result, leading to maximum delay improvement\\
\hline
\hline
\textbf{Input :} weight vector $\boldsymbol{\omega}=(\omega_1,\cdots,\omega_N)$, popularity matrix $\bold{P}$, caching indicator matrix $\mathbf{\Phi}$, source table $\bold{S}$, delay matrices $\bold{T_{\text{avg}}}$ and $\bold{D}$.\\
\textbf{Output :} new source table $\bold{S}$, new optimal delay matrix $\bold{D}$, and new caching indicator matrix $\mathbf{\Phi}$.\\
\hline
\textbf{Initialization :} set optimal delay improvement $g^*=0$, and set the corresponding $\bold{S}^*=\bold{S}$, $\bold{D}^*=\bold{D}$.\\
\\
\textbf{For} $i=1:N$\\
\hspace{0.5cm} \textbf{If} $\sum_{j=1}^{M} \phi_{i,j}<\mu$\\
\hspace{1cm} \textbf{For} $j=1:M$\\
\hspace{1.5cm} run Algorithm 1 for $<U_i,\mathcal{F}_j>$, to obtain\\
\hspace{1.5cm} the gain $g_{i,j}$ and the corresponding $\hat{\bold{S}}$ and $\hat{\bold{D}}$.\\
\hspace{1.5cm} \textbf{IF} $g_{i,j}>g^*$\\
\hspace{2cm} update $g^*=g_{i,j}$, $\bold{S}^*=\hat{\bold{S}}$, $\bold{D}^*=\hat{\bold{D}}$,\\
\hspace{2cm} $\widetilde{i}=i$, and $\widetilde{j}=j$.\\
\hspace{1.5cm} \textbf{End}\\
\hspace{1cm} \textbf{End}\\
\hspace{0.5cm} \textbf{End}\\
\textbf{End}\\
update $\phi_{\widetilde{i},\widetilde{j}}=1$, $\bold{S}=\bold{S}^*$ and $\bold{D}=\bold{D}^*$.
\\
\hline\hline
\end{tabular}
\end{footnotesize}
\end{table}

Based on Algorithm 1, Algorithm 2 described in Table \ref{Algorithm2} helps to find the optimal $<$file,user$>$ pair to be added to the updated caching result, which leads to the maximum delay reduction. In Algorithm 2, $\widetilde{i}$ and $\widetilde{j}$ record the optimal user index and file index, respectively. $g^*$ tracks the maximum delay improvement, and $\bold{S}^*$ and $\bold{D}^*$ record the new source table and minimum delay matrix, respectively, after caching $\mathcal{F}_{\widetilde{j}}$ at $U_{\widetilde{i}}$. We search over all $NM$ possible $<$file,user$>$ combinations, find their delay improvements and update $g^*$, $\widetilde{i}$, $\widetilde{j}$, $\bold{S}^*$ and $\bold{D}^*$ accordingly. At user $U_i$, we check if there is empty space in its cache. If its cache is full, we directly jump to the next user $U_{i+1}$. For each $<$file,user$>$ pair, we run Algorithm 1 to calculate the corresponding delay improvement, and compare it with $g^*$. If a $<$file,user$>$ pair exceeds the maximum delay improvement up to that point, we perform the update accordingly. Every time we run Algorithm 2, we cache one more file at a user. Therefore, we need to run Algorithm 2 $N\mu$ times to obtain the final caching result, and this process is described in Algorithm 3 in Table \ref{Algorithm3}.

\begin{table}
\caption{\label{Algorithm3}Algorithm 3}
\begin{footnotesize}
\begin{tabular}{p{8cm}}
\hline
\hline
Caching Algorithm\\
\hline
\hline
\textbf{Input :} weight vector $\boldsymbol{\omega}=(\omega_1,\cdots,\omega_N)$,  popularity matrix $\bold{P}$, and delay matrix $\bold{T_{\text{avg}}}$.\\
\textbf{Output :} caching indicator matrix $\mathbf{\Phi}$, source table $\bold{S}$.\\
\hline
\textbf{Initialization :} for all requests, $S_{i,j}\leftarrow \BS$, $D_{i,j}=T_{i,i}$. Set all $\phi_{i,j}=0$.\\
\textbf{For} $loop=1:N\mu$\\
\hspace{0.5cm} run Algorithm 2 to cache a file and update the result.\\
\textbf{End}\\
\hline\hline
\end{tabular}
\end{footnotesize}
\end{table}

For our proposed caching algorithm, we initially have all caches empty, and all users work in cellular mode, in which they only receive files from the base station at first. We assume that the system has calculated the average delay between every two nodes, and stored the delay matrix $\bold{T_{\text{avg}}}$ at the base station. Then, base station runs Algorithm 2 $N\mu$ times, and in each time we cache one more file and update the caching indicator $\mathbf{\Phi}$, source table $\bold{S}$, and minimum delay matrix $\bold{D}$ accordingly. Finally, the base station sends the caching files to the users when the traffic load is low.

\subsection{Complexity Analysis}

In the $l^{\text{th}}$ iteration, Algorithm 2 searches over $NM-(l-1)$ possible $<$file,user$>$ pairs, where the term $l-1$ corresponds to the $l-1$ $<$file,user$>$ pairs that have been selected in previous iterations. Therefore, the size of the search space of our algorithm is $\sum_{l=1}^{l=N\mu} NM-(l-1)=N^2M\mu-\frac{1}{2}N^2\mu^2+\frac{1}{2}N\mu$, which is much smaller than the size of the entire solution space $(\frac{M!}{(M-\mu)!\,\mu!})^N$.

In order to test the performance of our algorithm, we compare our algorithm with the brute-force exhaustive search algorithm. We apply both algorithms to a system, in which there are 5 users, 10 files in the library and each user can cache 2 files. These two algorithms obtain the same caching result, however the time consumption of the exhaustive search algorithm is $1.28\times 10^5$ seconds, while our algorithm only takes only $2.7\times 10^{-3}$ seconds.

\section{Extensions and Future Work}\label{sec:extension}
In this section, we consider a more general case, in which the delay matrices $\bold{T_{\text{avg}}}$ and $\bold{D}$, weight vector $\boldsymbol{\omega}=(\omega_1,\cdots,\omega_N)$, popularity matrix $\bold{P}$ and transmission powers $P_i$ change over time. For simplicity, we assume that all these parameters stay constant within one update cycle, and we use $\kappa$ as the index of cycles. The duration of the $\kappa^{\text{th}}$ cycle, denoted by $\tau^\kappa$, depends on how fast the parameters vary. Then, we can formulate our caching problem in the $\kappa^{\text{th}}$ cycle as
\begin{small}
\begin{align}
\textbf{P2:}\hspace{0.5cm}&\text{Minimize}_{\;\mathbf{\Phi}^{\kappa}}\hspace{0.5cm} \sum_{i=1}^{N} \omega_i^{\kappa} \sum_{j=1}^{M} P_{i,j}^{\kappa}D_{i,j}^{\kappa}\\
&\text{Subject to} \hspace{0.7cm} \sum_{j=1}^{M} \phi_{i,j}^{\kappa}=\mu\\
&\hspace{2.2cm} \sum_{j=1}^{M} \left|\phi_{i,j}^{\kappa}-\phi_{i,j}^{\kappa-1}\right|\leq 2\xi_i^{\kappa}\label{constraint_2}\\
&\hspace{2.2cm} \phi_{i,j}^{\kappa}\in\{0,1\}.
\end{align}
\end{small}
If we define the weight of user $i$ as $\omega_i^\kappa=\E\{\text{NPK}_i^\kappa/\text{NPK}^\kappa\}$, where $\text{NPK}_i^\kappa$ and $\text{NPK}^\kappa$ represent the number of received packets in the $\kappa^{\text{th}}$ cycle at user $i$ and at all users, respectively, then the objective function in the optimization problem \textbf{P2} represents the expected packet delay in the $\kappa^{\text{th}}$ cycle. The transmission power $P_i^\kappa$ is determined according to the battery budget of user $i$. Due to the changes in transmission powers and the distributions of channel fading, the delay matrices $\bold{T_{\text{avg}}}^\kappa$ and $\bold{D}^\kappa$ also vary over time. Compared with \textbf{P1}, \textbf{P2} includes an additional constraint given by (\ref{constraint_2}). In (\ref{constraint_2}), $\xi_i^{\kappa}$ is the upper bound of the number of cache files that will be replaced in the current update cycle. Due to requirements regarding energy efficiency and current traffic load, each user may be able to update only a few cache contents.

The solution of \textbf{P2} is described below:
\begin{enumerate}
  \item At the beginning of the $\kappa^{\text{th}}$ cycle, the system estimates the delay matrix $\bold{T_{\text{avg}}}^{\kappa-1}$, weight vector $\boldsymbol{\omega}^{\kappa-1}$, and popularity matrix $\bold{P}^{\kappa-1}$ according to the samples obtained in the previous cycle. The base station receives the transmission powers $P_i^\kappa$ from the users, determine the cycle period $\tau^\kappa$ and the upper bound $\xi_i^{\kappa}$, and then predicts $\bold{T_{\text{avg}}}^{\kappa}$, $\boldsymbol{\omega}^{\kappa}$ and $\bold{P}^{\kappa}$.
  \item Algorithm 2 is repeated $N\mu$ times to determine the caching result in the $\kappa^{\text{th}}$ cycle.
  \item At the end of each iteration in the second step, it is checked if the constraint in (\ref{constraint_2}) is satisfied with equality at any one of the users. If this constraint is satisfied with equality at a user, then no more cache updating is allowed for this user, meaning that this user can only choose from the files that are already stored in its cache in the remaining iterations.
\end{enumerate}
After this process, the base station sends the cache contents to each user, and conduct regular transmission after updating the cache files at each user.

As we have mentioned in Section \ref{Sec:formulation}, this improved algorithm does not require the first $4$ resource allocation assumptions described in Section \ref{sec:model}, and works for any resource allocation algorithm, since the delay matrices $\bold{T_{\text{avg}}}$ and $\bold{D}$ need to be evaluated via estimation or learning methods. Also, we note that this method requires estimation algorithms in the first step. Due to the page limitations, we leave a detailed study of this problem as our future work.

\section{Numerical Results}\label{Sec:numerical}
In this section, we investigate the performance of our proposed algorithm via numerical results. Since the estimation and resource allocation components required for the extended algorithm in Section \ref{sec:extension} are beyond the scope of this paper, we only consider Algorithm 3 and its corresponding system model in this section. In the numerical results, the location of each user is randomly generated within a circular cell with the base station placed at the center. Each point in the figures is obtained by taking average over $500$ randomly generated systems. The popularity matrix is generated according to the Zipf distribution. When the users have identical popularity, they give the same popularity index to a file, which leads to identical rows in the popularity matrix $P$. When the users have independent popularity, each user gives popularity indices to the files independently. In other words, identical popularity indicates that all users have the same preference, while independent popularity indicates that each user has an independent preference. The number of files in the library is $M=100$, and the size of each file is $11.3$ bits. We assume Rayleigh fading with path loss $\E\{z\}=d^{-4}$, where $d$ represents the distance between the transmitter and the receiver. The transmission powers are set as $P_b=23\:dB$ and $P_u=20\:dB$, and we choose the weights as $\omega_i=\hat{p}_i$ so that $\eta$ represents the average system delay.

\begin{figure*}
\hspace{-1cm}
\begin{minipage}[b]{0.35\linewidth}
\begin{center}
\includegraphics[width=\textwidth]{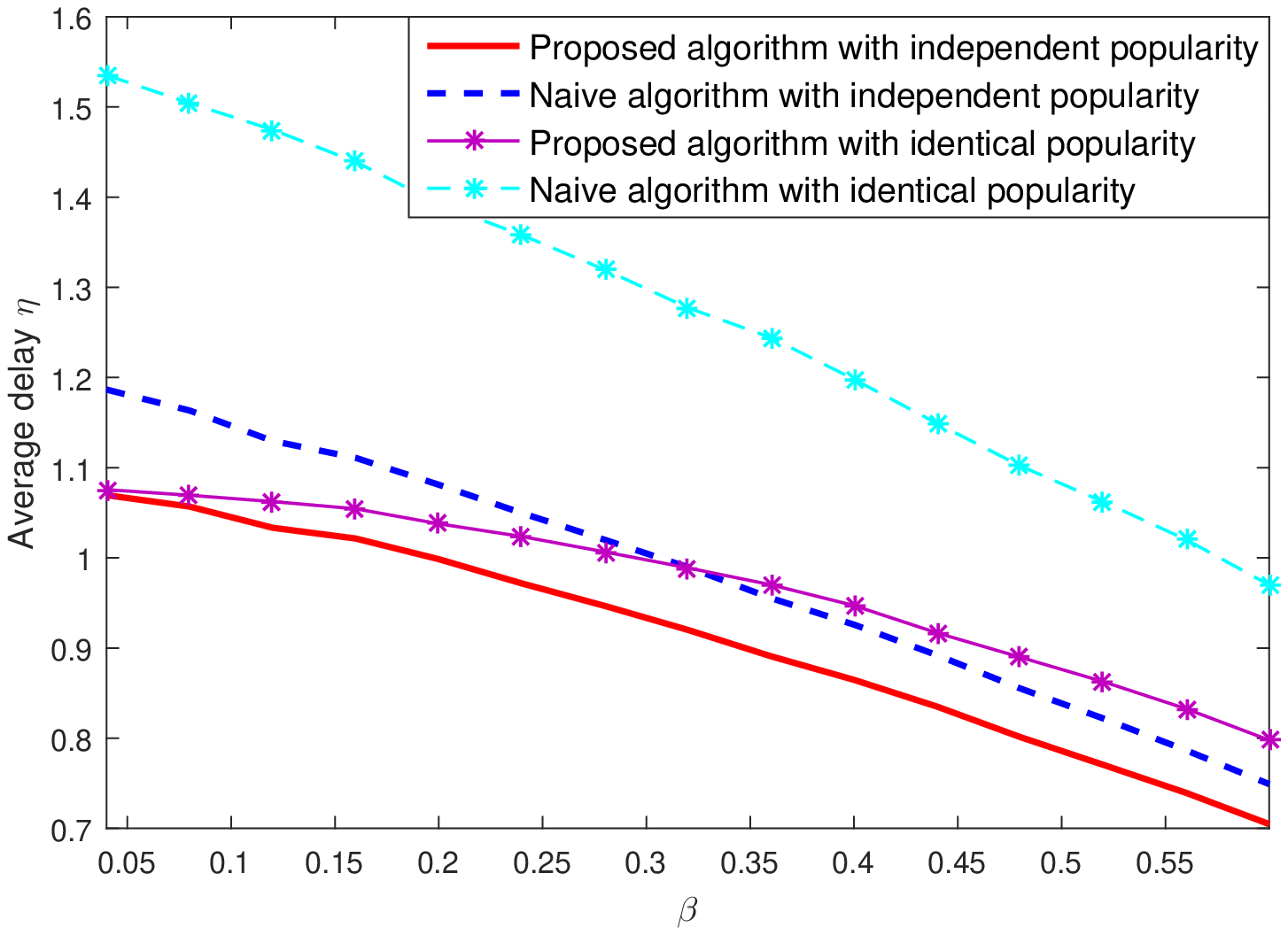}
\caption{Average delay $\eta$ vs. Zipf exponent $\beta$}\label{fig:fig1}
\end{center}
\end{minipage}
\begin{minipage}[b]{0.35\linewidth}
\begin{center}
\includegraphics[width=\textwidth]{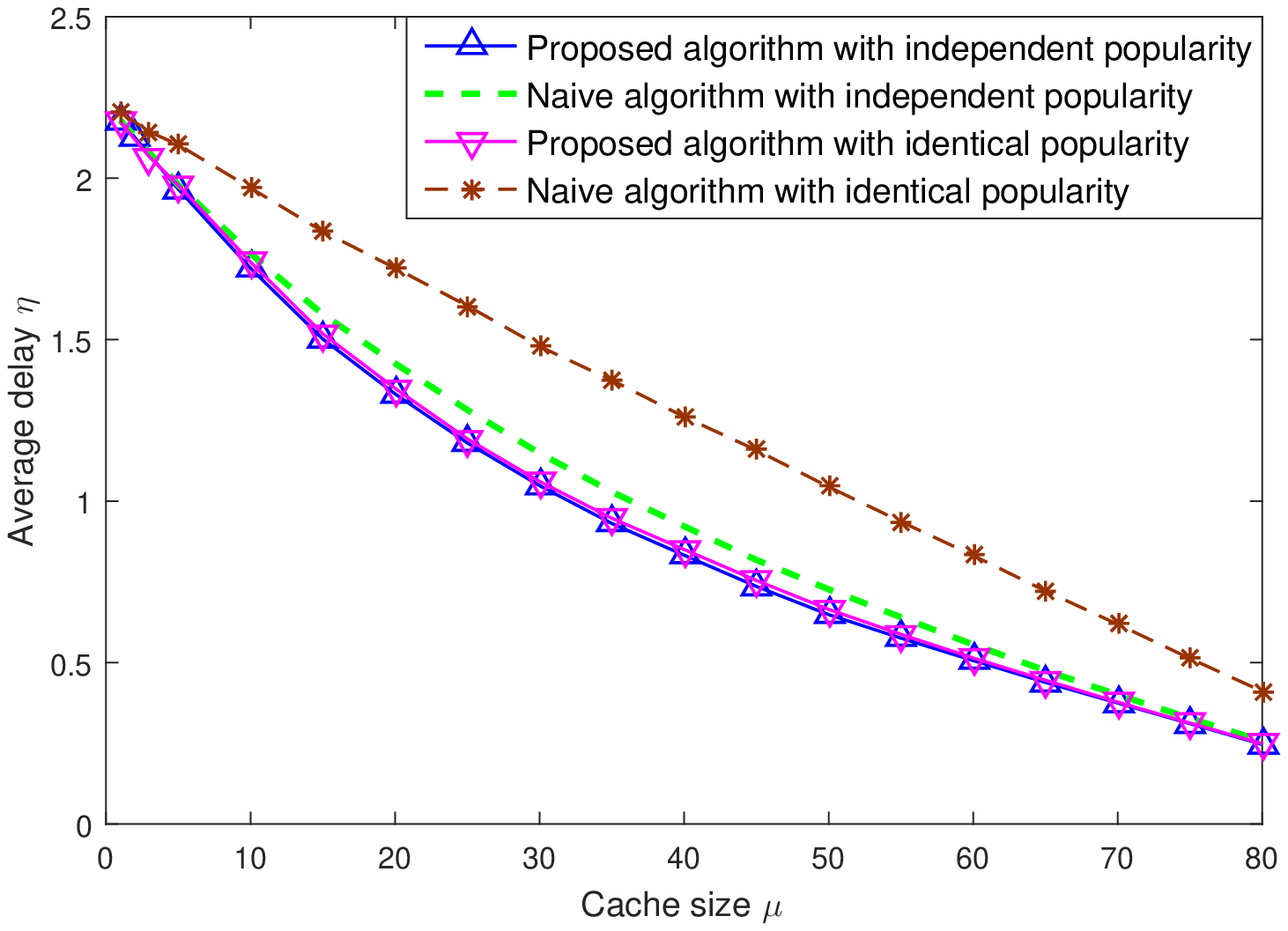}
\caption{Average delay $\eta$ vs. cache size $\mu$}\label{fig:fig2}
\end{center}
\end{minipage}
\begin{minipage}[b]{0.35\linewidth}
\begin{center}
\includegraphics[width=\textwidth]{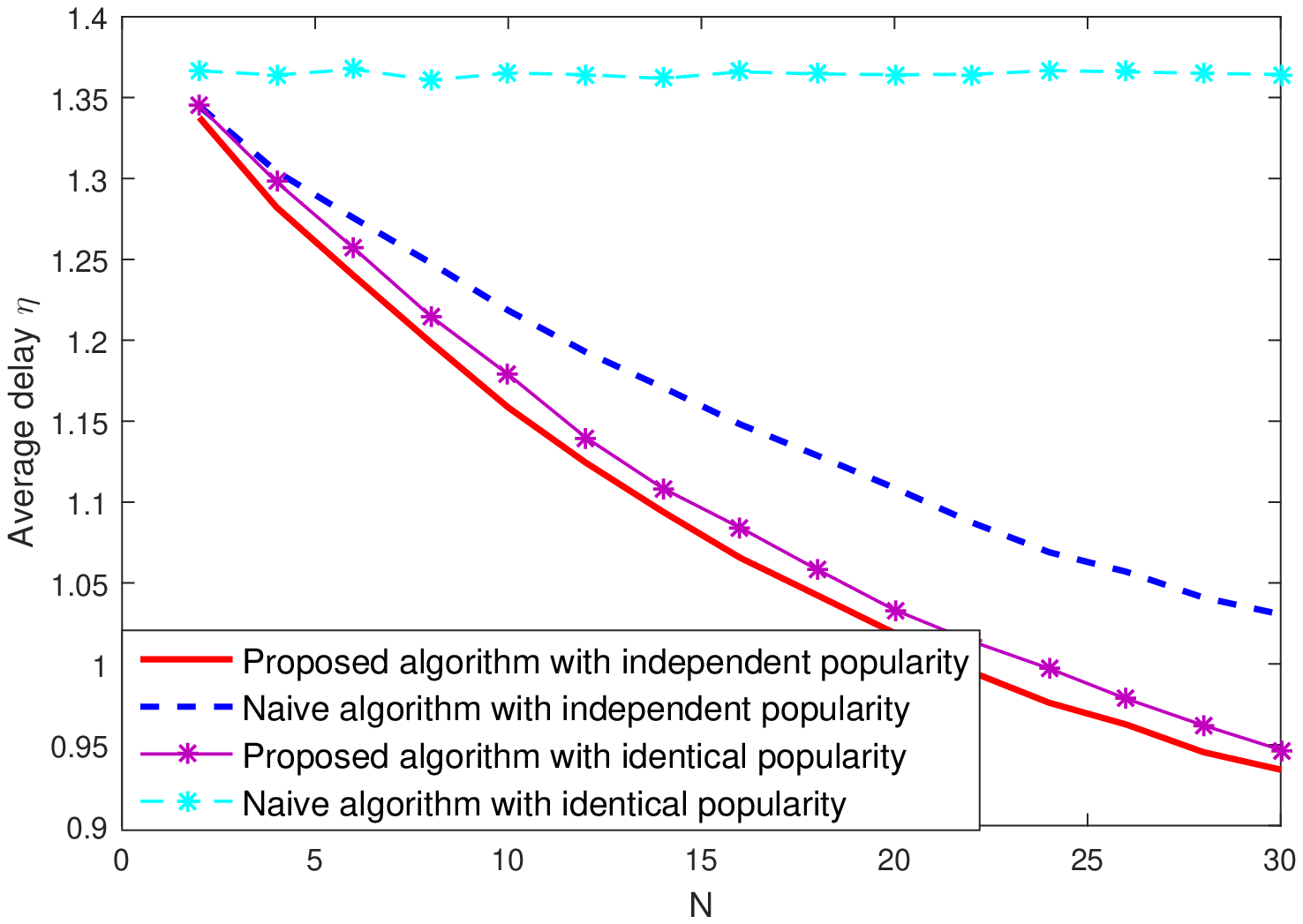}
\caption{Average delay $\eta$ vs. the number of users $N$}\label{fig:fig3}
\end{center}
\end{minipage}
\end{figure*}

In the numerical results, we compare the performance of our proposed algorithm with a naive algorithm, in which each user just caches the most popular $\mu$ files. This naive algorithm is efficient when the base station does not have the knowledge of the channel fading statistics and the cached files at each user. In this circumstance, the users just cache files according to their own preference. In the case of naive algorithm with identical popularity, every user caches the same files, and they get the files they do not have via cellular downlink from the base station. Therefore, the gap between the two curves using naive algorithm in Figs. \ref{fig:fig1}-\ref{fig:fig3} (which will be discussed in detail next) demonstrates the benefit of enabling D2D communications. By allowing D2D transmission, the users far away from the base station can get files from their neighbors, which helps to significantly reduce the delay.

In Fig. \ref{fig:fig1}, we set $N=25$, $\mu=30$ and plot the average delay $\eta$ as a function of the Zipf exponent $\beta$. As $\beta$ increases, the popularity difference increases. When $\beta=0$, the users request all files with equal probability; when $\beta\rightarrow +\infty$, each user only requests its most favorite file. Therefore, we only need to concentrate on the delay performance of fewer popular files as $\beta$ increases, and it becomes easier to achieve better delay performance with limited caching space. That is the reason for having monotonically decreasing curves in Fig. \ref{fig:fig1}. Another observation is that our algorithm is more robust to the popularity setting. Compared to the curves using the naive algorithm, identical popularity model only slightly raises the delay of our algorithm. If a node can get a popular file from its near neighbor, then caching some less popular files might give better delay improvement. Therefore, our algorithm can enable D2D transmission even in an identical popularity model, which guarantees the robustness.


In Fig. \ref{fig:fig2}, we select $\beta=0.1$, $N=25$ and plot the average delay as a function of the caching size $\mu$. When $\mu$ is small, the delay difference between different algorithms and different popularity settings is small. In such a situation, both algorithms cache the most popular files. As $\mu$ increases, the difference in performance increases. As we have mentioned in Algorithm 2, our algorithm searches for the optimal $<$file,user$>$ pair that provides the maximum delay improvement, and this mechanism guarantees a very sharp decrease at the beginning. After exceeding a threshold, further increasing the caching size reduces the performance difference, because the system gets enough caching size to cache most of the popular files. Overall, Fig. \ref{fig:fig2} shows that our algorithm can achieve better delay performance with limited caching size.


In Fig. \ref{fig:fig3}, we select $\beta=0.1$, $\mu=30$ and plot the average delay as a function of the number of users $N$. For the curve using the naive algorithm with identical popularity model, having more users does not affect the average delay because each user works in cellular mode and receives the files from the base station. For other curves, increased number of users enables more chances for D2D communication, and as a result the average delay decreases. Compared with the naive algorithm, our algorithm can achieve better performance when the number of users is large.


\section{Conclusion}
In this paper, we have proposed a caching algorithm for D2D cellular networks, which minimizes the weighted average delay. First, we have characterized the popularity model and average transmission delay of a request. Then, we have formulated the delay minimization problem and developed our algorithm which can solve the weighted average delay minimization problem efficiently. We have also extended our algorithm for a more general scenario, in which the distributions of fading coefficients and system parameters change over time. Finally, we have further investigated the performance of our algorithm by comparing it with a naive algorithm which simply caches the most popular files at each user. By applying both algorithms to two different popularity models, we have shown that our algorithm is more robust to variations in the popularity models, and can achieve better performance, because the proposed algorithm can more effectively take advantage of D2D communications. Also, the influence of the popularity parameter, caching size and number of users is studied via numerical results.

\bibliographystyle{ieeetr}
\bibliography{D2D_caching}

\end{document}